\documentstyle[aps,twocolumn,epsfig]{revtex}
\begin{document}
\newcommand{\ve}[1]{\mbox{\boldmath $#1$}}
\twocolumn[\hsize\textwidth\columnwidth\hsize
\csname@twocolumnfalse%
\endcsname

\draft

\title {Phases of a rotating Bose-Einstein condensate with
anharmonic confinement}
\author{A. D. Jackson$^1$, G. M. Kavoulakis$^2$, and E. Lundh$^3$}
\date{January 14, 2004}
\address{$^1$Niels Bohr Institute, Blegdamsvej 17, DK-2100 Copenhagen \O,
             Denmark \\
         $^2$Mathematical Physics, Lund Institute of Technology, P.O.  Box
             118, SE-22100 Lund, Sweden \\
     $^3$Department of Physics, KTH, SE-10691, Stockholm, Sweden}
\maketitle

\begin{abstract}

We examine an effectively repulsive Bose-Einstein condensate of atoms
that rotates in a quadratic-plus-quartic potential. With use of a variational
method we identify the three possible phases of the system (multiple
quantization, single quantization, and a mixed phase) as a function of the
rotational frequency of the gas and of the coupling constant. The derived
phase diagram is shown to be universal and the continuous transitions 
to be exact in the limit of weak coupling and small anharmonicity. 
The variational results are found to be
consistent with numerical solutions of the Gross-Pitaevskii equation.

\end{abstract}
\pacs{PACS numbers: 03.75.Hh, 03.75.Kk, 67.40.Vs}

\vskip0.5pc]

\section{Introduction}
When rotated, superfluids are expected to show a structure of
quantized vortex states.  This effect has been confirmed
experimentally in Bose-Einstein condensates of atoms which
typically are confined in harmonic potentials
\cite{JILA,Madison,VortexLatticeBEC,HaljanCornell,JILAgiant}.
In contrast to traditional experiments on superfluid liquid helium
contained in a ``bucket'', the presence of a harmonic potential
introduces the trap frequency, $\omega$, which sets the scale
for the critical frequency of rotation of the gas $\Omega_c$ for
vortex formation.  The same frequency $\omega$ also sets an upper
bound for the rotational frequency of the atoms.  When $\Omega = \omega$,
the centrifugal force exactly cancels the confining force, and the
atoms fly apart.  In harmonically trapped Bose-Einstein
condensates, the cloud starts to rotate when $\Omega = \Omega_c$, and
a vortex state forms at its center.  As $\Omega$ increases, more
vortices appear and eventually form an array.  As indicated, $\Omega$ cannot
exceed the trap frequency, $\omega$.

Theoretical studies suggest that this picture becomes richer when an
anharmonic term is added to the trapping potential
\cite{Lundh,Fetter,tku,FG,KB,Emil,SPB,AA,JK}. The presence
of the anharmonic term leads to two significant differences.
First, $\Omega$ is no longer bounded by $\omega$.  Second, the
system can exhibit multiply-quantized vortex states, i.e., vortex
states with more than one units of circulation (as opposed to
singly-quantized vortex states which have one unit of circulation.)
Indeed, for fixed interaction strength and sufficiently large $\Omega$,
the gas always exists in a state of multiple quantization.  In the opposite
extreme of fixed $\Omega$ and large interaction strength, the gas forms
an array of vortices as in the case of harmonic confinement.
In the intermediate regime where both parameters are large, there is a
third phase which contains both an array of vortices and a strong
suppression of the density in the middle of the cloud.  This state
is a mixed phase with a multiply-quantized vortex state at the center of
the cloud and singly-quantized vortices around it.
Recently, the experiment of Ref.\,\cite{Dal} managed to create
vortices in an anharmonic trapping potential and investigated the
phases of the system as a function of $\Omega$.  Reference \cite{Dal2} has
examined the dynamics of the gas in the presence of vortices.

The purpose of the present study is to offer a description of the
transitions between phases of multiple and single quantization. These
transitions can be of either first or second order.  In the limit of
weak coupling and small anharmonicity, this will lead to an exact
description of the continuous (i.e., second-order) transitions.
In the same limit, the phase diagram as a function of
the strength of the interaction versus $\Omega$ will be seen to be
universal. This phase diagram turns out to be very rich. Of particular
importance are the triple points which we discover, which indicate the
meeting of one phase of multiple quantization with two other phases
which are distinct combinations of singly- and multiply-quantized
vortex states with discrete rotational symmetries. In fact, we encounter
an entire family of triple points.

We present our model in Sec.\ \ref{sec:model}.  We then examine separately 
continuous phase transitions in Sec.\ \ref{sec:continuous} and discontinuous 
phase transitions in Sec.\ \ref{sec:discontinuous}. In Sec.\ \ref{sec:universality},
we demonstrate the universality of the phase diagram
and the location of the phase boundaries describing continuous transitions
in the limit of small anharmonicity and weak coupling.  We offer a physical
interpretation of our results in Sec.\ \ref{sec:physical}.
Section \ref{sec:exact} contains the phase
diagram that results from the solution of the exact eigenvalue problem for the
anharmonic confining potential and Sec.\ \ref{sec:numerical} discusses the 
direct numerical solutions of the Gross-Pitaevskii equation, which are in
agreement with our variational approach.  Finally Sec.\ \ref{sec:conclusions} 
gives a summary of our results and conclusions.

\section{Model -- Approach}
\label{sec:model}

We start by assuming a confining potential of the form
\begin{equation}
   V(\rho) = \frac 1 2 M \omega^2 \rho^2 [1 + \lambda (\frac {\rho} {a_0})^2].
\label{anh}
\end{equation}
Here $\rho$ is the cylindrical polar coordinate, $M$ is the atomic mass,
$a_0 = (\hbar/M \omega)^{1/2}$ is the oscillator length, and $\lambda$
is a small dimensionless constant.  In the experiment of Ref.\,\cite{Dal},
$\lambda \approx 10^{-3}$.  We neglect the trapping potential along the
$z$ axis since the cloud rotates about this direction and instead assume
a constant density per unit length $\sigma = N/Z$, where $N$ is the number
of atoms and $Z$ is the width of the gas along the $z$ axis. The energy
levels of the non-interacting (two-dimensional) problem are thus given by
\begin{eqnarray}
  \epsilon = (2 n_r + |m| + 1) \hbar \omega,
\label{energy}
\end{eqnarray}
where $n_r$ is the radial quantum number, and $m$ is the quantum number
corresponding to the angular momentum.  This equation emphasizes the
large degeneracy which is associated with the many different ways of
distributing $L$ units of angular momentum to $N$ atoms.  As shown in
Refs.\,\cite{Rokhsar,KMP}, this degeneracy is lifted by the interactions.
One of the basic conclusions of these studies is that repulsive interactions
always favor singly-quantized vortex states.  On the other hand, when
$\lambda > 0$ and the interaction is sufficiently weak, it is energetically
favorable for the system to form multiply-quantized vortex states.  As
the interaction increases, the effective repulsion between vortices
ultimately results in the splitting of multiply-quantized vortex states
into singly-quantized states \cite{Lundh,Fetter,tku,FG,KB,Emil,SPB,AA,JK}.

Here we shall account for the interactions with the usual assumption of
a contact potential
\begin{eqnarray}
   V_{\rm int} =
   \frac 1 2 U_{0} \sum_{i \neq j} \delta({\bf r}_{i} - {\bf r}_{j}),
   \label{v}
\end{eqnarray}
where $U_0 = 4 \pi \hbar^2 a/M$ is the strength of the effective
two-body interaction with $a$ equal to the scattering length for
atom-atom collisions. The interaction is assumed repulsive,
$a > 0$. (Reference \cite{Emil} has examined an effectively-attractive
Bose-Einstein condensate confined in an anharmonic potential.)
The dimensionless quantity which plays the role of a coupling constant
in our two-dimensional problem is thus $\sigma a$, since the typical
atom density $n$ is $\sim N/(\pi a_0^2 Z)$ in the limit of weak interactions
that we consider here.  Therefore, the typical interaction energy $n U_0$
is $\sim \sigma a \hbar \omega$.

As shown in Ref.\,\cite{JK}, the phase transition between multiple
and single quantization occurs when $\sigma a$ is $\approx \lambda/\alpha$,
where $\alpha$ is a dimensionless constant of order $10^{-1}$. Given
that $\lambda \approx 10^{-3}$ in Ref.\,\cite{Dal}, $\sigma a$ is less than
unity in the region of the transition. This fact allows us to adopt a
variational approach where the unperturbed states are restricted to
the nodeless eigenstates of the harmonic potential with angular momentum
$m \hbar$
\begin{equation}
    \Phi_m({\rho, \phi}) = \frac 1 {(m! \pi a_0^2 Z)^{1/2}}
       \left( \frac {\rho}{a_0} \right)^{|m|} e^{i m \phi}
        e^{-\rho^{2}/2 a_0^2}.
\label{phim}
\end{equation}
Here $\phi$ is the angle in cylindrical polar coordinates.
In this basis we can write the order parameter as
\begin{eqnarray}
   \Psi(\rho, \phi) = \sum_m c_{m} \Phi_{m},
\label{exptrm}
\end{eqnarray}
where the coefficients $c_{m}$ are variational parameters (assumed to be
real without loss of generality).  As a first step, we calculate the
energy. From Ref.\,\cite{JK} we find that the energy per particle in the
rotating frame in the state $\Psi$ given by Eq.\,(\ref{exptrm}) is
\begin{eqnarray}
  {\cal E} = \frac E N =
     \frac 1 S \sum_m \epsilon_m c_m^2
\phantom{XXXXXXXXXXX} \nonumber \\
      + \frac 1 {S^2}
        \sum_{m,n,l,k} c_m c_n c_l c_k \langle m,n | V_{\rm int} | l,k \rangle
     \delta_{m+n,l+k},
\label{engen}
\end{eqnarray}
where $S = \sum_m c_m^2$ and
\begin{eqnarray}
        \frac {\epsilon_m} {\hbar \omega}
      = |m| -  m \frac {\Omega} {\omega} + \frac \lambda 2 (|m|+1)
      (|m|+2)
\label{sen}
\end{eqnarray}
is the single-particle energy in the rotating frame measured with respect
to the zero-point energy, $\hbar \omega$.  Here the matrix elements of
the interaction are given as
\begin{eqnarray}
     \langle m, n | V_{\rm int} | l, k \rangle =
    \sigma a \frac {(|m| + |n|)!}
     {2^{|m| + |n|} \sqrt{|m|! |n|! |l|! |k|!}} \delta_{m+n,l+k}.
\label{intdme}
\end{eqnarray}

As shown in Ref.\,\cite{JK}, for sufficiently small $\sigma a$ this energy
is minimized when only one component $m = m_0$ in the above sum is nonzero,
i.e., $c_{m} = 0$ for $m \neq m_0$.  The critical frequencies, which denote
the lower limit on the absolute stability of the state $m_0$, are given
by \cite{JK}
\begin{eqnarray}
   \frac {\Omega_{m_0}} {\omega} = 
   \frac {m_0} {|m_0|} \left[ 1 + \lambda (|m_0|+1)
\right. \phantom{XXXXX} \nonumber \\ \left.
         - \sigma a \frac {(2 |m_0| - 2)!} {2^{2 |m_0| - 1} (|m_0|-1)! |m_0|!}
     \right] .
\label{gen}
\end{eqnarray}
The (straight) solid lines in Fig.\,1 show the phase boundaries in the
$\Omega/\omega$ -- $\sigma a$ plane as given by Eq.\,(\ref{gen}). The
transitions across these phase boundaries are discontinuous and of first
order.

As $\sigma a$ increases, however, the states with a single component $m = m_0$
become unstable resulting in both continuous as well as discontinuous phase
transitions. In Sec.\,III we identify the phase boundaries within a given
sector $m_0$ as continuous transitions which are denoted as dashed lines in
Fig.\,1.  Clearly, so long as $\sigma a$ is smaller than the value for which
a given solid line is first cut by a dashed line, the straight solid lines
correctly locate the discontinuous phase transitions between
multiply-quantized vortices.  On the other hand, for somewhat larger values of
$\sigma a$ there is a competition between the mixed state that lies to
the right of each solid line and the multiply-quantized state to its left.
As a result of this competition, the straight lines of Eq.\,(\ref{gen})
no longer describe the phase boundary correctly; the correct boundary is
pushed to the left as shown in the dotted-dashed lines in Fig.\,1.  The
determination of this boundary is given in Sec.\ \ref{sec:discontinuous}.
\noindent
\begin{figure}
\begin{center}
\epsfig{file=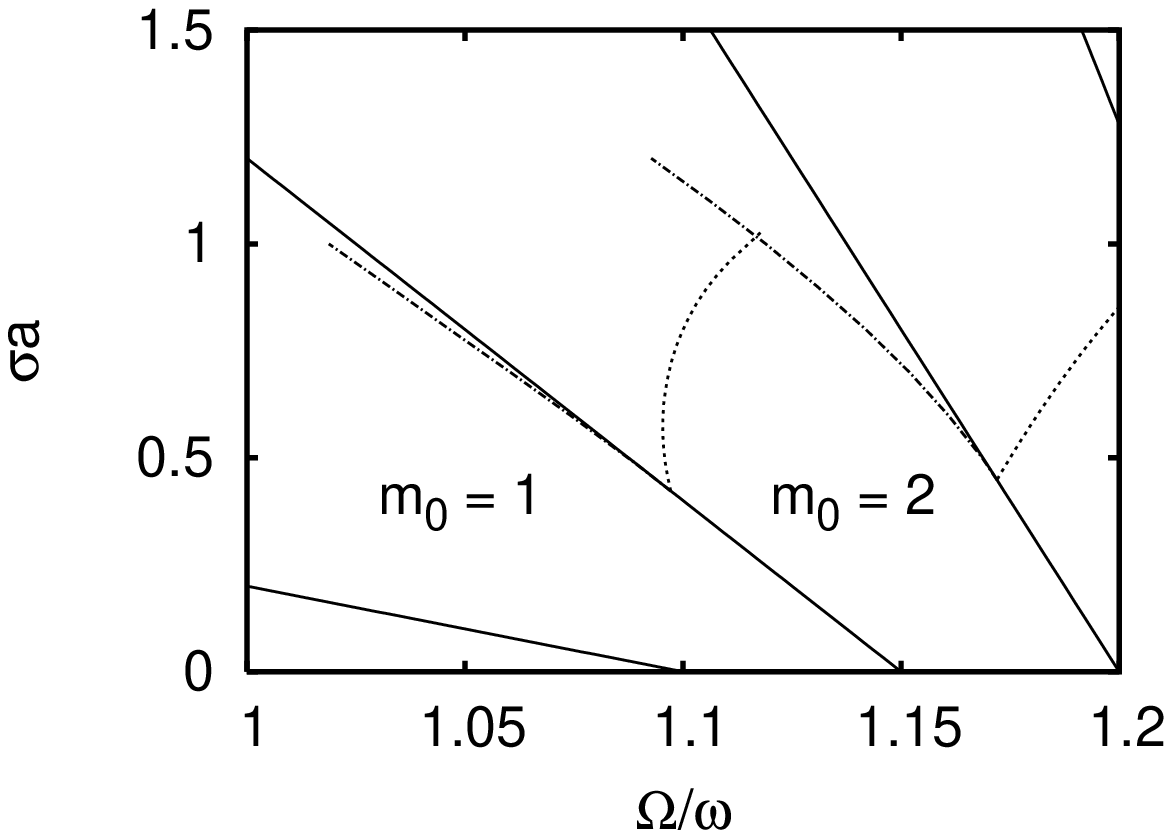,width=9.0cm,height=7.0cm,angle=0}
\vskip0.5pc
\begin{caption}
{The phase diagram in the $\Omega/\omega$ -- $\sigma a$ plane for $\lambda
= 0.05$. The straight lines show the phase boundary between multiply-quantized
vortices given by Eq.\,(\ref{gen}), which denote discontinuous transitions. The
dotted lines give the phase boundary between $m_0 = 2$ and $(m_1, m_0, m_2) =
(0, 2, 4)$ (left), and between $m_0 = 3$ and $(m_1, m_0, m_2) = (0, 3, 6)$ 
(right), which denote continuous transitions. The dotted-dashed lines show the
phase boundary between $m_0 = 1$ and $(m_1, m_0, m_2) = (0, 2, 4)$ (left),
and between $m_0 = 2$ and $(0, 3, 6)$ (right), which denote discontinuous
transitions. The straight lines do not have any physical meaning for
$\sigma a$ higher than the point where they are cut by the dashed lines,
as the phase boundary is given by the dotted-dashed lines.}
\end{caption}
\end{center}
\label{FIG1}
\end{figure}

\section{Continuous transitions}
\label{sec:continuous}

We begin our consideration of the continuous transitions by showing that
along the phase boundary, where only the $m = m_0$ state is present (and
thus $c_{m} = 0$ for $ m \neq m_0 $), the derivative of the energy with
respect to all the $c_{m}$ vanishes.  Having established that this pure
state is an extremum, we will then examine the matrix that results by
calculating the second derivatives of the energy with respect to any
$c_{m}$ and $c_{n}$. The criterion for the stability of the pure state
$m_0$ is then the positivity of the eigenvalues of this matrix.  An
instability and, hence, the phase boundary occurs when one eigenvalue of
this matrix becomes negative.

It is important to note that close to the phase boundary one needs
to keep only those terms in the interaction energy [i.e., in the last term
of Eq.\,(\ref{engen})] which are at most bilinear in the $c_m$ for $m
\neq m_0$. This implies that the only two terms which must be retained
are those with $\langle m_0, m_i | V_{\rm int} | m_0, m_i \rangle$
($i = 1, 2$) and $\langle m_0, m_0 | V_{\rm int} | m_1, m_2 \rangle$ with
$m_1 + m_2 = 2 m_0$.  The second term is the only off-diagonal element
in the second-derivative matrix.  This observation simplifies the problem
significantly since the matrix of second derivatives is block diagonal with
the 
\noindent
\begin{figure}
\begin{center}
\epsfig{file=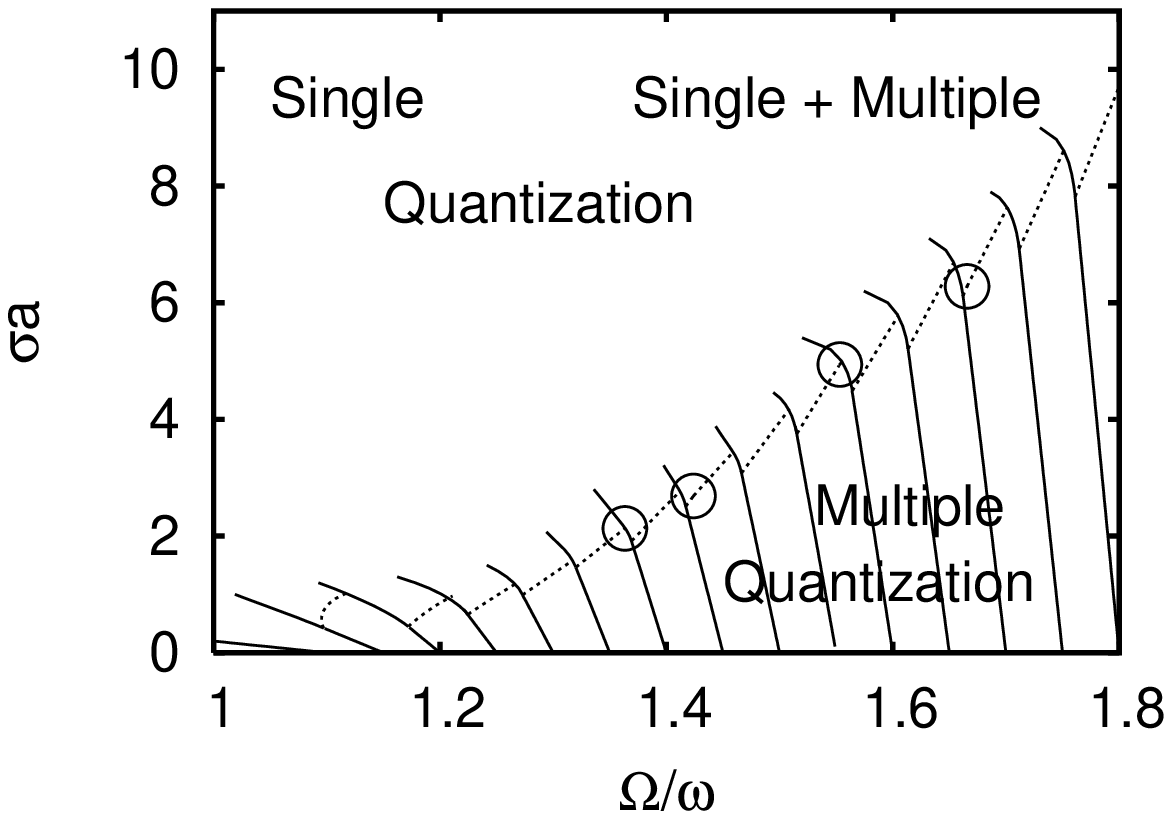,width=9.0cm,height=7.0cm,angle=0}
\vskip0.5pc
\begin{caption} 
{The phase diagram of vortex states in a quadratic-plus-quartic potential
in the $\Omega/\omega$ -- $\sigma a$ plane for $\lambda = 0.05$ with
$m_0=0$,\dots,15. The solid lines denote discontinuous transitions, and the
dashed lines continuous transitions. The first circle on the left indicates
the triple point, and the other circles show triple points on the lines of
continuous transitions for $m_0 =8$, 10, 13.}
\end{caption}
\end{center}
\label{FIG2}
\end{figure}
\noindent
dimensionality of each block being equal to two at most. The problem
reduces to diagonalizing 2 $\times$ 2 matrices.

Given the form of the interaction, Eq.\,(\ref{intdme}), it is elementary
to verify that all derivatives $\partial E / \partial c_m$ do indeed vanish
at the phase boundary as a consequence of angular momentum conservation
[i.e., the delta function in Eq.\,(\ref{intdme})].  Turning to the
second derivatives, $\partial^2 E / \partial c_m \partial c_n$, explicit
calculation reveals that $\partial^2 E / \partial c_{m_0}^2 = 0$ and that
\begin{eqnarray}
   \frac {\partial^2 {\cal E}} {\partial c_{m_i}^2}
   = 2 \hbar \omega (\epsilon_{m_i} - \epsilon_{m_0})
\phantom{XXXXXXXXX} \nonumber \\
   + [8 \langle m_0, m_i | V_{\rm int} | m_0, m_i \rangle
      - 4 \langle m_0, m_0 | V_{\rm int} | m_0, m_0 \rangle],
\label{dme}
\end{eqnarray}
and
\begin{eqnarray}
   \frac {\partial^2 {\cal E}} {\partial c_{m_i} \partial c_{m_j}} =
     4 \langle m_0, m_0 | V_{\rm int} | m_1, m_2 \rangle
     \delta_{2m_0,m_1+m_2}.
\label{senme}
\end{eqnarray}
Using Eqs.\,(\ref{intdme}), (\ref{dme}), and (\ref{senme}), we can now
construct and diagonalize the resulting matrix, which is block diagonal
as a consequence of the delta function in Eq.\,(\ref{senme}).
Evidently, the highest dimensionality of each block is 2 $\times$ 2. The
result is shown as the dashed curves in Fig.\,2 for $m_0 \le 15$. For
$m_0 \leq 6$, we find that the most unstable mode is that which
involves the states with $(m_1, m_0, m_2) = (0, m_0, 2m_0)$.  However,
as $\Omega$ and $m_0$ increase, it becomes energetically unfavorable to
have a significant atom density close to the center of the cloud.
This is due to the form of the effective potential
\noindent
\begin{figure}
\begin{center}
\epsfig{file=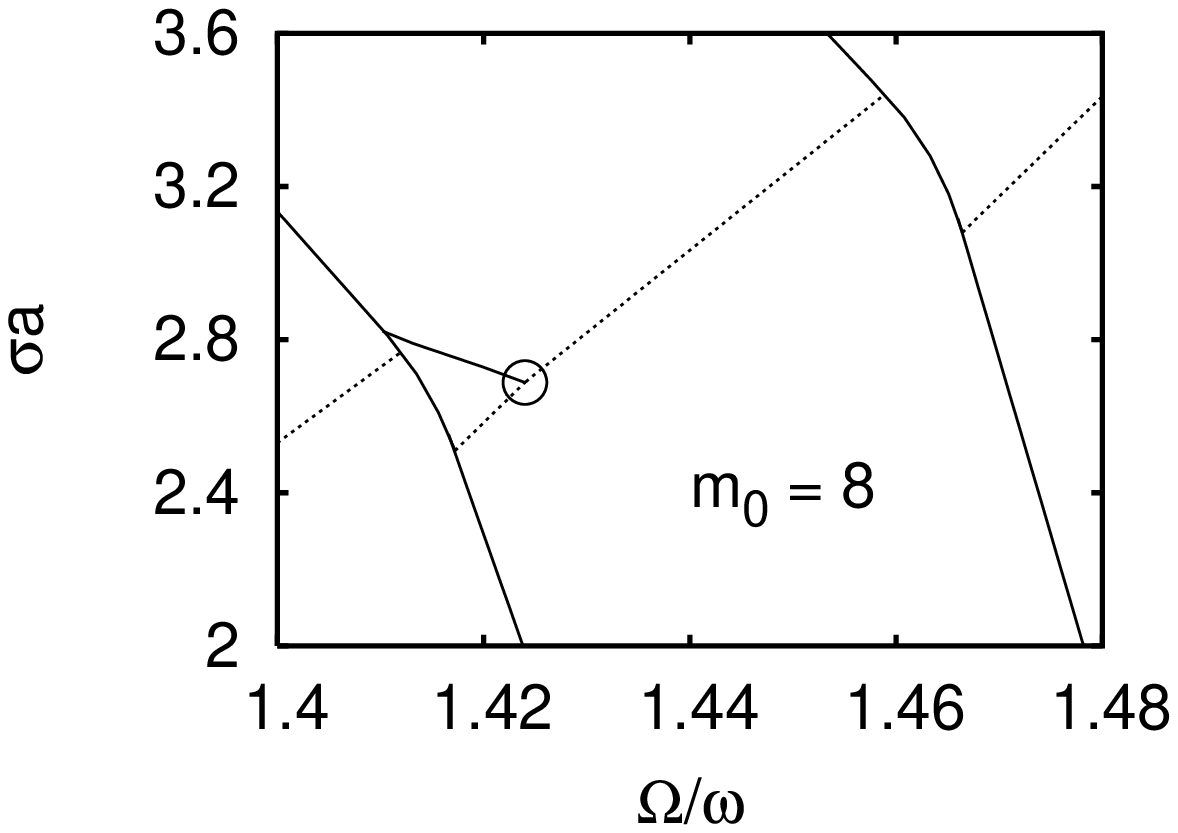,width=9.0cm,height=7.0cm,angle=0}
\vskip0.5pc
\begin{caption}
{The same graph as in Fig.\,2 around the phase boundary for $m_0 =8$.
Solid lines denote discontinuous transitions and dashed lines continuous.
The slope of the dashed curve changes at the center of the circle, which
is a triple point. In the small triangular region the state is
dominated by the $(m_1, m_0, m_2) = (2, 8, 14)$ components, while 
above the circle it is dominated by the $(1, 8, 15)$ components.
On the left the states are $m_0 = 7$ (bottom), and $(1, 7, 13)$ (top).
On the right the states are $m_0 = 9$ (bottom), and (2, 9, 16) (top).}
\end{caption}
\end{center}
\label{FIG3}
\end{figure}
\begin{eqnarray}
     V_{\rm eff}(\rho) = \frac 1 2 M (\omega^2 - \Omega^2) \rho^2
           + \frac {\lambda} 2 M \omega^2 \frac {{\rho}^4} {{a_0}^2},
\label{effp}
\end{eqnarray}
which has the familiar ``Mexican hat'' shape for $\Omega / \omega > 1$.
Since $\Phi_{0}$ is the only state which does not vanish at the origin,
this implies that the admixtures of $\Phi_{0}$ to the ground state
wave function will become less favorable as $\Omega/\omega$ increases.
Ultimately, the leading instability will involve states with $m_1 \ne 0$,
and the ground state will have a hole at its center with a density that
vanishes like $\rho^{2 m_1}$.  Our results show that this happens at
$m_0 = 7$ where the most unstable mode involves the states $(m_1, m_0, m_2)
= (1, 7, 13)$ instead of (0, 7, 14).  These states lead to the most unstable
mode over the entire region defined by Eq.\,(\ref{gen}).

In the region where the $m_0 = 8$ mode is stable, the picture becomes
even more interesting.  Initially, the most unstable mode involves the
states with $(m_1, m_0, m_2) = (2, 8, 14)$.  When $\Omega/\omega = 1.4240$
and $\sigma a = 2.6880$, however, the state with $(1, 8, 15)$ becomes
the most unstable.  This implies that there is a genuine triple point
involving the pure multiply-quantized vortex state with $m_0 = 8$ and
two phases which both have a vanishing density at the origin but which have
different symmetries. As we shall see, the three phases that meet at 
the triple point correspond to a single multiply-quantized vortex, a 
lattice of singly quantized vortices, and a multiply quantized
vortex surrounded by singly quantized ones.
Figure 3 shows these results. The circle denotes
this genuine triple point.  Note that the slope of the phase boundary
(i.e., the dashed line) changes at this point.

For higher values of $m_0$ this pattern persists with transitions
from some $m_0$ to either one set of $(m_1, m_0, m_2)$ extending over the whole
region of stability of the $m_0$ state or with a transition from $m_0$
to one set of states $(m_1, m_0, m_2)$ initially, and then to some other set
$(m_1', m_0, m_2')$.  The second alternative is realized for $m_0 = 10$
[which is initially unstable to $(3, 10, 17)$ and finally to $(2, 10, 18)$]
and for $m_0 = 13$ [which is initially unstable to $(5, 13, 21)$ and finally
to $(4, 13, 22)$].  These triple points are indicated by the final two
circles in Fig.\,2.  For $m_0 \le 15$, these are the only two patterns
found.  Note that we cannot exclude the possibility of more triple
points along a given segment of the second-order line for larger values of
$m_0$.  In addition, $m_1$ is found to be an increasing function of $m_0$.
We offer a physical interpretation of these results in
Sec.\ \ref{sec:physical}.

\section{Discontinuous transitions}
\label{sec:discontinuous}

Turning to the discontinuous transitions, it is instructive to consider a
specific example between the sectors with $m_0 = 1$ and $m_0 = 2$
(see Fig.\,1).  The value of $\sigma a$ at which the first dashed line cuts
the solid line is 0.4226.  For $\sigma a < 0.4226$ the solid line thus
gives the correct phase boundary between the pure states with $m_0 = 1$ and
$m_0 = 2$, and the transition is discontinuous.  For $\sigma a > 0.4226$,
there is a competition between the pure state $\Psi=\Phi_1$ and the state
$\Psi= c_0 \Phi_0 + c_2 \Phi_2 + c_4 \Phi_4$ identified in Sec.\,III (to
the left of the dashed line).  Comparison of the energy of the system in the
rotating frame in these two states allows us to find the boundary between the
two phases (dotted-dashed line). Since each state describes a local energy
minimum, the transition between them is still discontinuous and of first
order. Beyond $\sigma a = 0.4226$ the solid line is thus meaningless, and the
dotted-dashed line determines the phase boundary.  Note that the slope
of the solid line is constant while that of the dotted-dashed line
is not.  This implies a discontinuity in the curvature at the joining of
these curves.  The use of a better variational wave function on the right
of the solid curve indicates that the correct phase boundary must be
pushed to the left.  This is clearly seen in the results of Figs.\,1, 2, 3,
and 5.

As mentioned above, the triple point, i.e., the point where the three
phases coexist \cite{KB}, lies on the boundary between the states with
$m_0 = 6$, $(m_1, m_0, m_2) = (0, 6, 12)$, and $(1, 7, 13)$.  To determine
its precise location we need to combine the
calculations presented in Secs.\,III and IV.  This triple point is found to
be at $\Omega/\omega = 1.3633$ and $\sigma a = 2.1300$ and is indicated
by the first circle in the phase diagram of Fig.\,2.

\section{Universality and exactness of the phase boundary}
\label{sec:universality}

The phase diagram of Fig.\,2 is universal in the limit of weak coupling
and small anharmonicity.  More precisely, if given values of $\sigma a$
and $1-\Omega/\omega$ locate a point on the phase boundary of Fig.\,2
for a specific $\lambda$, the corresponding values become $(\sigma a)' =
\beta \sigma a$ and $(1-\Omega/\omega)' = \beta (1-\Omega/\omega)$ for
another $\lambda' = \beta \lambda$.  As $\lambda$ changes, both axes
scale by the same amount and in that sense the phase diagram of Fig.\,2 is
universal.  Of course, this conclusion is valid only if the anharmonicity
and the interaction are both sufficiently weak to justify our restriction
to radial states with $n_r = 0$.

The method adopted in the present study of the continuous transitions is
also exact in the limit of weak coupling and small anharmonicity since,
infinitesimally close to the corresponding boundaries, one can neglect all
terms with $m \neq m_0$ and $m \neq m_i$ in the expansion of
Eq.\,(\ref{exptrm}).  (In this regard, see also Sec.\,VI.)  As one
proceeds away from the phase boundary, additional angular momentum
states must of course be included.  With increasing coupling but arbitrarily
close to the phase boundary, it may be necessary to consider states with
$n_r \neq 0$.  However, the restrictions on the angular momentum can be
maintained.  Thus, our approach can be used to explore the phase boundary
for significantly larger values of $\Omega / \omega$ and $\sigma a$ by
including additional radially excited states as needed.  The phase boundary
between multiply-quantized vortex states is also calculated exactly in the
same limit close to the line of second-order transitions.  On the other hand,
the calculation of phase boundaries for the discontinuous transitions
involving multiply-quantized vortices and the mixed phases (i.e., the case
considered in Sec.\,IV) is only approximate since additional angular
momentum states are in principle required for the description of the mixed
phase.

\section{Physical interpretation of the results}
\label{sec:physical}

It is useful to examine the spatial distribution of vortices and their
multiplicity. The order parameter in the phases of multiple quantization
is $\Psi = \Phi_{m_0}$, and just above the phase boundaries it has the
form
\begin{equation}
   \Psi = (c_{m_1} N_{m_1} {\tilde z}^{m_1}
           + c_{m_0} N_{m_0} {\tilde z}^{m_0}
       + c_{m_2} N_{m_2} {\tilde z}^{m_2})
   e^{-|\tilde z|^{2}/2 a_0^2},
\label{svsi}
\end{equation}
where $m_1 < m_0 < m_2$ and ${\tilde z} = \rho e^{i \phi} = x + i y$.
A numerical calculation reveals that close to the phase boundary and
to leading order,
\begin{equation}
   c_{m_1} \propto c_{m_2} \propto \delta^{1/2}, \,\,\,\,
   c_{m_0} - 1 \propto \delta^{1/2},
\label{svsi2}
\end{equation}
where $\delta = \sigma a - (\sigma a)_c$.  Here $(\sigma a)_c$ is the
value of $\sigma a$ on the phase boundary for a fixed $\Omega$.
It is straightforward 
\begin{figure}
\begin{center}
\epsfig{file=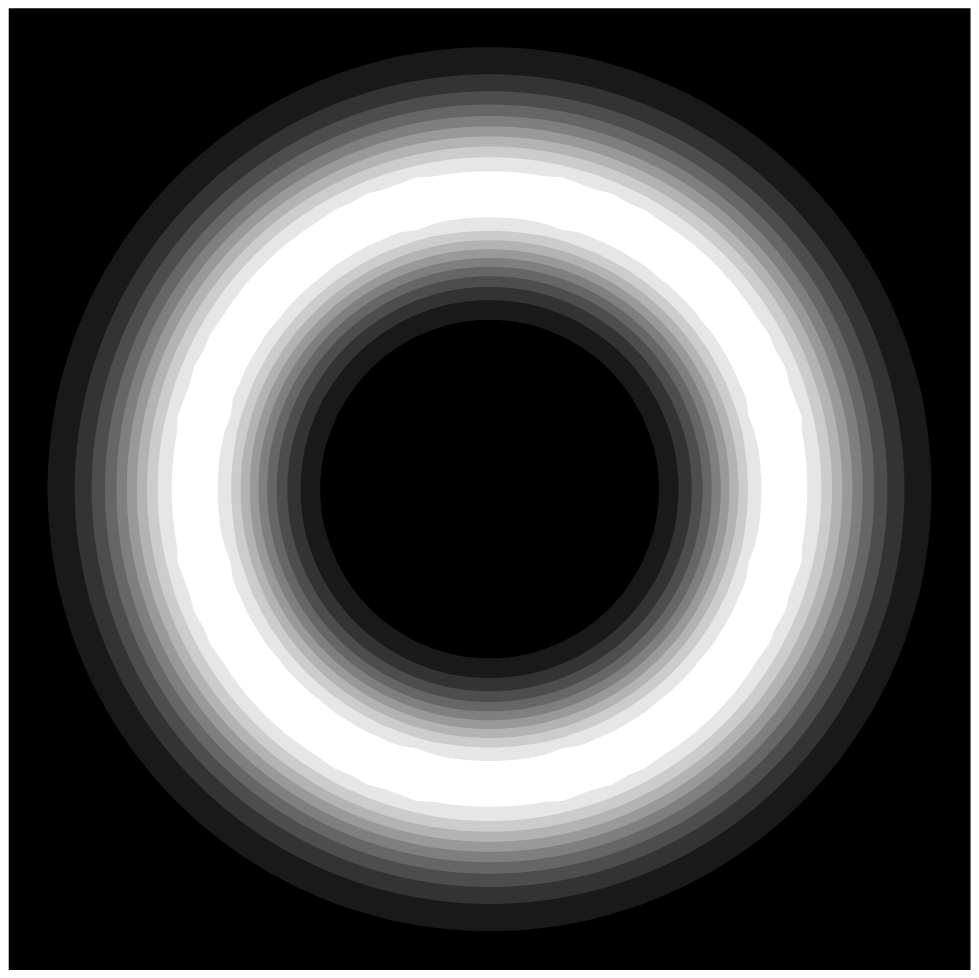,width=2.0cm,height=2.0cm,angle=0}
\epsfig{file=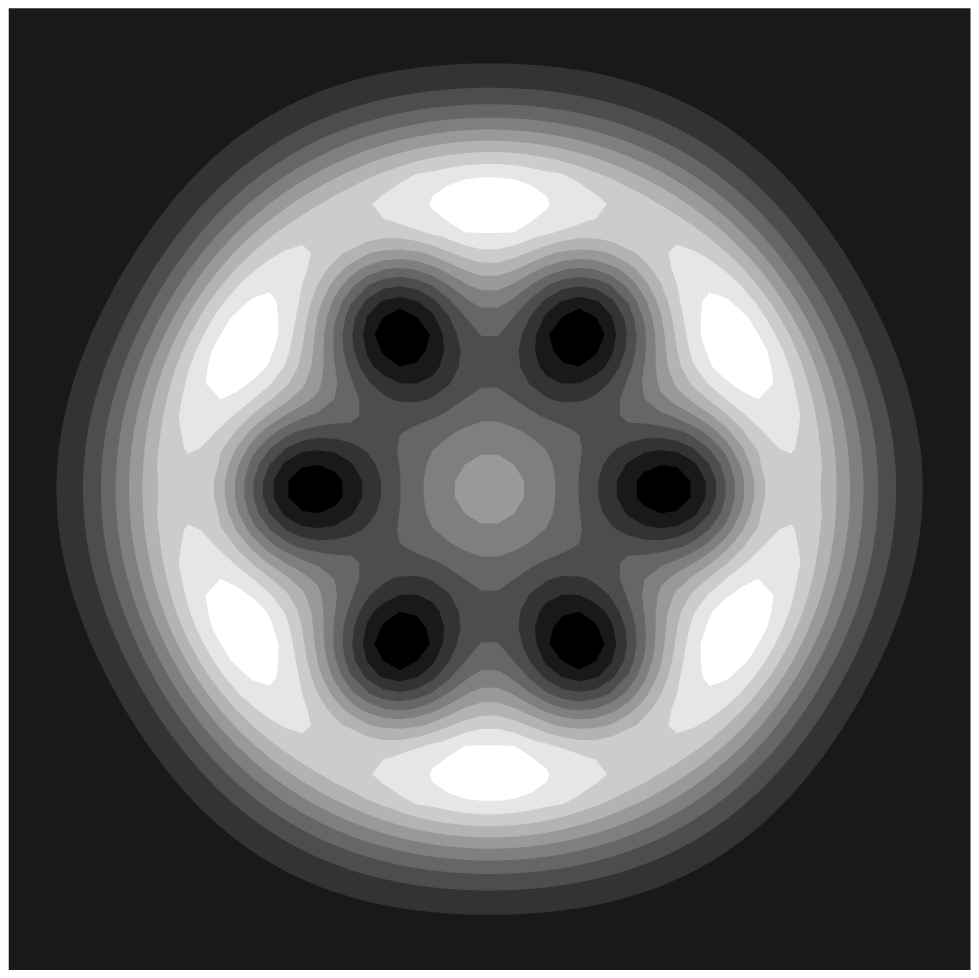,width=2.0cm,height=2.0cm,angle=0}
\end{center}
\begin{center}
\epsfig{file=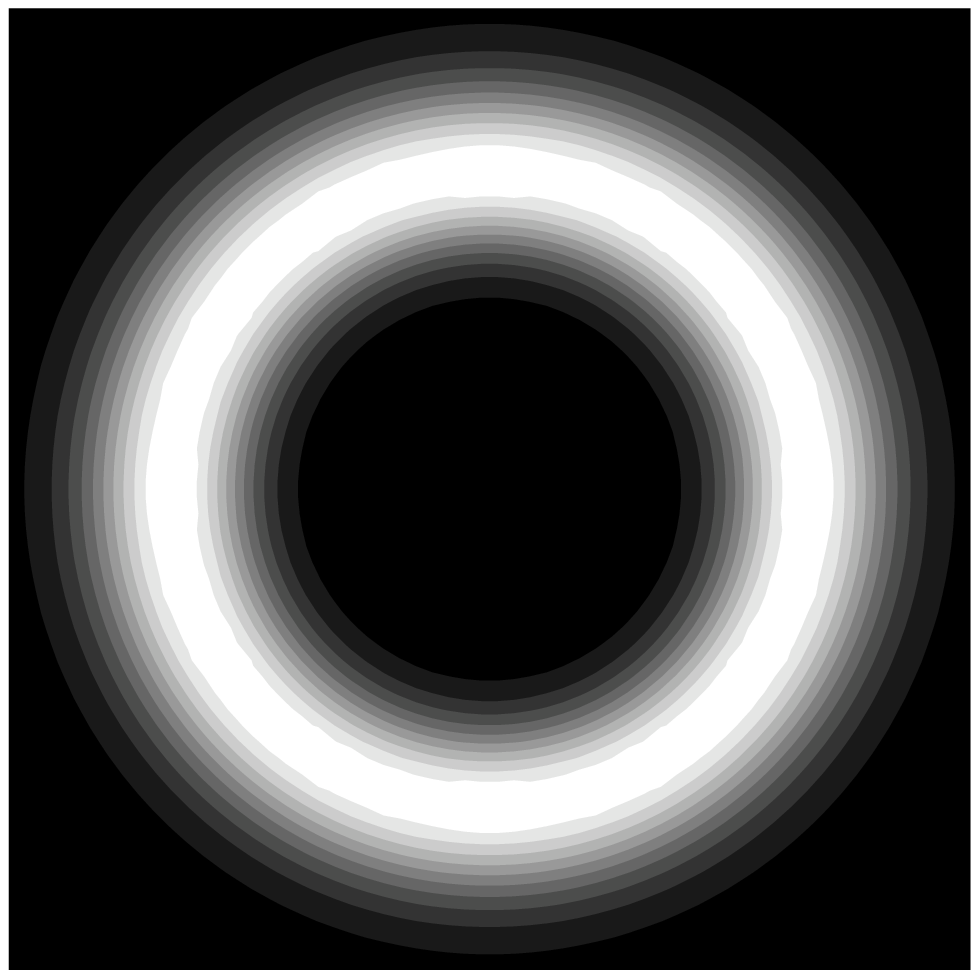,width=2.0cm,height=2.0cm,angle=0}
\epsfig{file=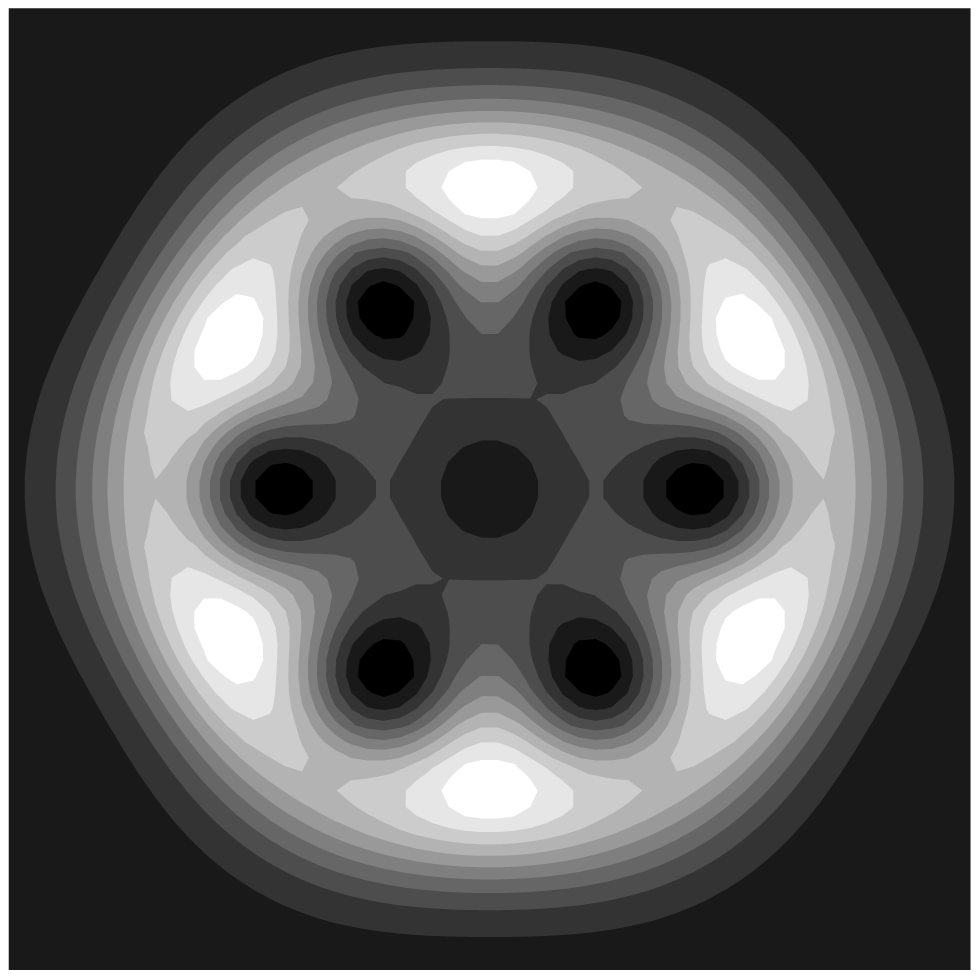,width=2.0cm,height=2.0cm,angle=0}
\end{center}
\begin{center}
\epsfig{file=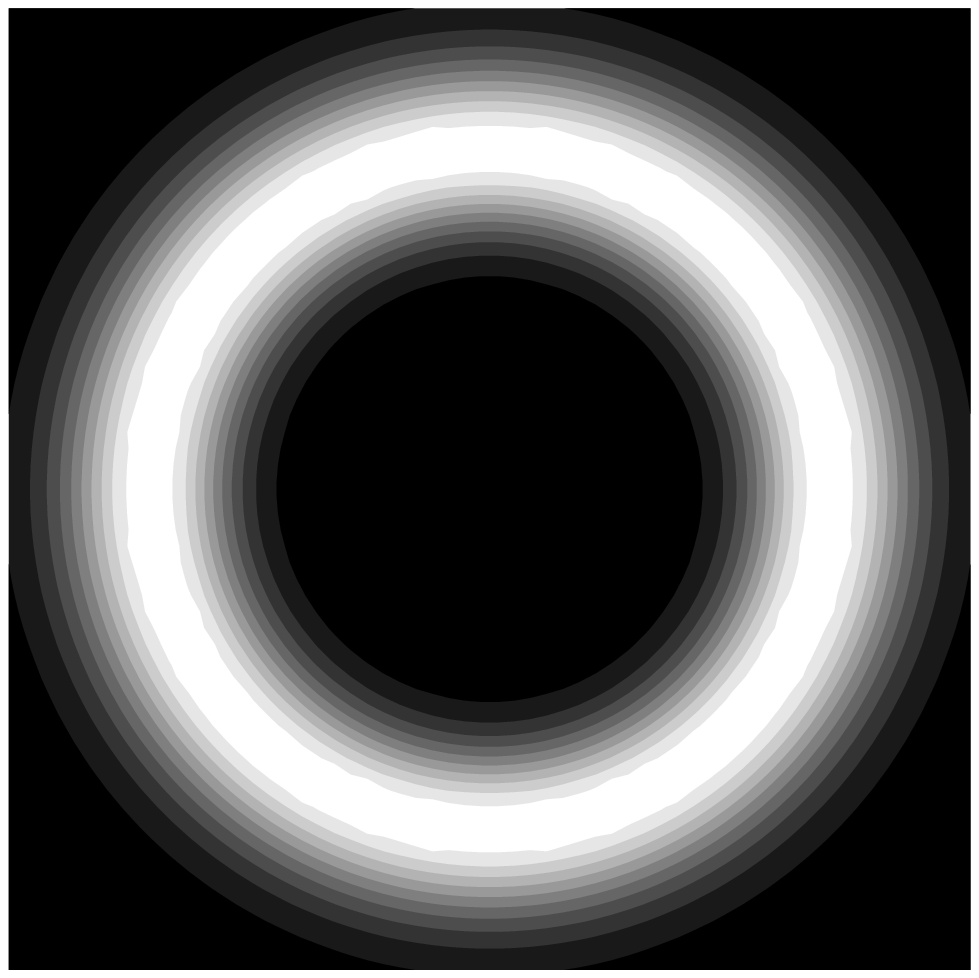,width=2.0cm,height=2.0cm,angle=0}
\epsfig{file=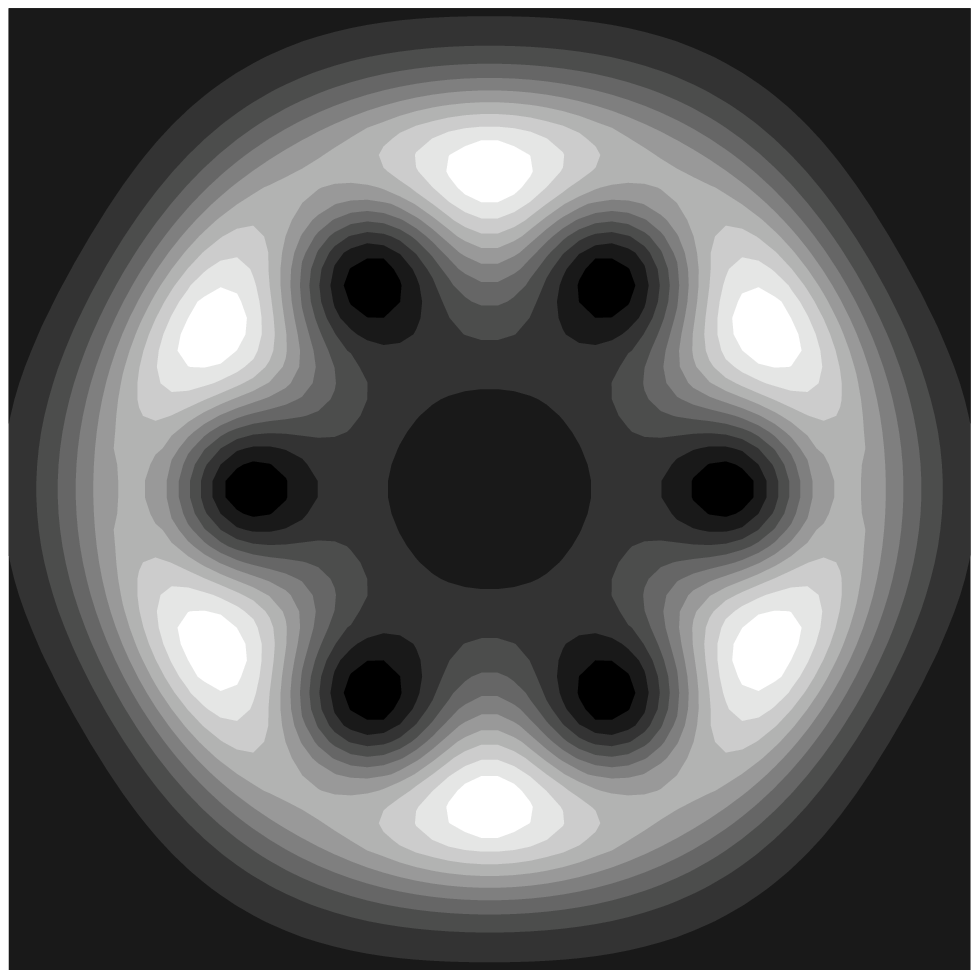,width=2.0cm,height=2.0cm,angle=0}
\epsfig{file=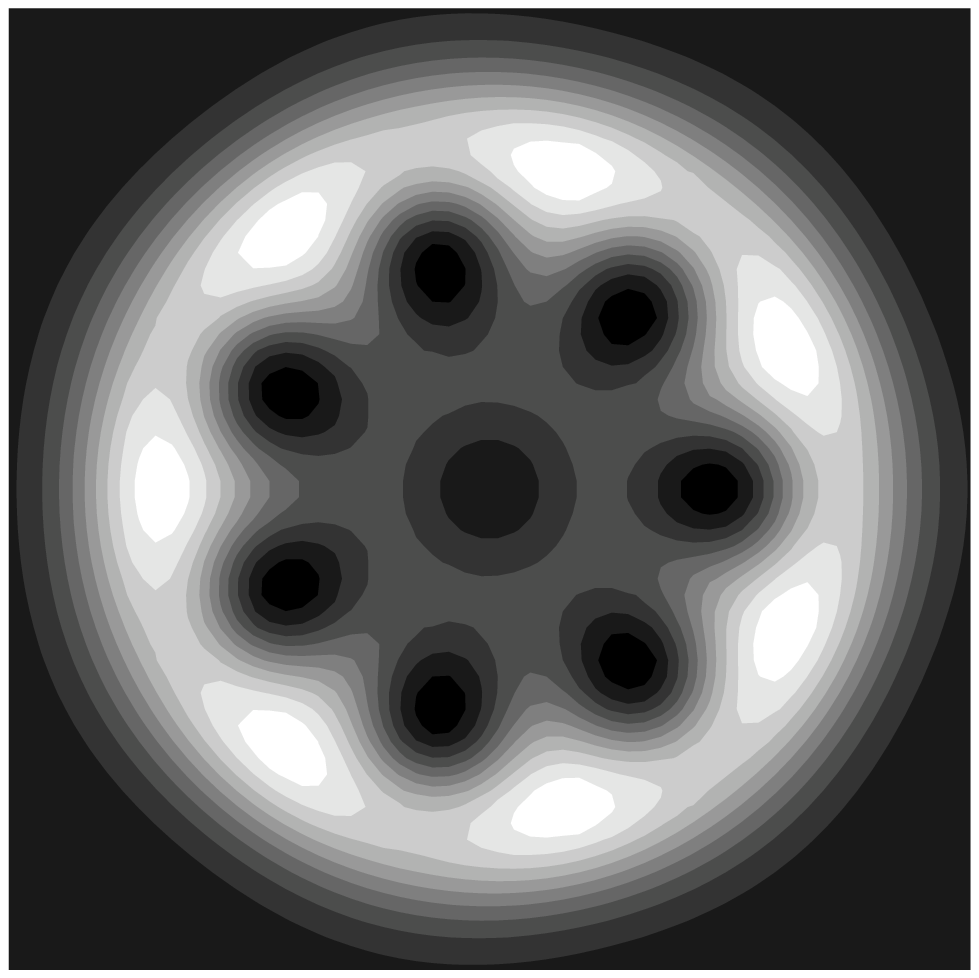,width=2.0cm,height=2.0cm,angle=0}
\vskip0.5pc
\begin{caption}
{Contour plots of the density of the cloud. The width in both axes
is from $-4 a_0$ up to $4 a_0$. The white color denotes high density
and the black low density. The upper figures show the density of the
cloud in the state with only $m_0 = 6$ (left) and $(m_1, m_0, m_2) =
(0, 6, 12)$ (right). The middle figures refer to the case with only
$m_0 = 7$ (left) and $(1, 7, 13)$ (right). The bottom figures refer
to the case with only $m_0 = 8$ (left), $(2, 8, 14)$ (middle), and
$(1, 8, 15)$ (right). For $m_0 \ge 7$ there is always a node in
the density at the center of the cloud.}
\end{caption}
\end{center}
\label{FIG4}
\end{figure}
\noindent
to develop a power-series expansion in $\delta$
for all coefficients that are not strictly zero (as a consequence of angular
momentum conservation) \cite{KMP}. For example, the next coefficient $c_m$
contributing to the state of Eq.\,(\ref{svsi}) is determined by the condition
$c_{m_0}^2 c_{m}^2 \sim c_{m_1} c_{m_0} c_{m_2} c_{m}$ where
$m = m_0 + m_2 - m_1$.  This implies that
\begin{equation}
  c_m \propto c_{m_1} c_{m_2} \propto \delta,
\end{equation}
The next coefficient $c_{m'}$ is determined by
$c_{m_0}^2 c_{m'}^2 \sim c_{m_0} c_{m_2} c_{m} c_{m'}$, where
$m' = m_2 + m - m_0 = 2 m_2 - m_1$, and thus
\begin{equation}
  c_{m'} \propto c_{m_2} c_{m} \propto \delta^{3/2}.
\end{equation}
Evidently, the higher coefficients can be neglected at the phase boundary.

Returning to Eq.\,(\ref{svsi}), $\Psi$ has $m_2$ nodes in the $xy$
plane, and each of them corresponds to a vortex state.  However, only
the $m_0$ of them are located in a region of non-negligible density
since the density drops rapidly away from the center of the cloud due
to the exponential factor. More precisely, because of the first term
on the right of Eq.\,(\ref{svsi}), at a small $\delta$ above the
phase boundary there is still a multiply-quantized vortex state at the
center of the cloud with ${m_1}$ units of circulation.  In addition,
there are $m_0 - m_1$ simple zeros or singly-quantized vortices surrounding
it at a distance $R_1$ which is
\begin{equation}
    \frac {R_1} {a_0} \sim
   \left( \frac {\delta m_0!} {m_1!} \right)^{1/[2(m_0 - m_1)]}.
\end{equation}
The remaining vortices are located at a distance $R_2$
\begin{equation}
   \frac {R_2} {a_0} \sim
   \left( \frac {m_0!} {\delta m_2!} \right)^{1/[2(m_2 - m_0)]}.
\end{equation}

For example, when the $m_0 = 8$ state becomes unstable to $m_1 = 2$,
$m_0 = 8$, and $m_{2} = 14$, there is a doubly-quantized vortex at the
center of the gas, six more vortices around it at a distance of order
$R_1 /a_0 \sim 2.3 \delta^{1/12}$ and a final six at a distance of order
$R_2/a_0 \sim 3.4/ \delta^{1/12}$, where the density is essentially zero
for sufficiently small $\delta$ (see the bottom graphs in Fig.\,4). Since
$m_1$ increases with increasing $m_0$, just above the phase boundary a
multiply-quantized vortex state splits into a multiply-quantized
vortex state with lower circulation, with the remaining (singly-quantized)
vortices surrounding it. This is a mixed phase, and the hole which
develops in the middle of the cloud is a multiply-quantized vortex
state with a quantum number $m_1$, in agreement with Refs.\cite{Lundh,tku}.
It is worth mentioning that the maximum of the function $\rho |\Phi_m|^2$
occurs for $\rho/a_0 = \sqrt{m + 1}$, which gives the typical size of
a vortex state with $m$ quanta.

Figure 4 shows contour plots of the density of the cloud for $m_0 = 6, 7$,
and 8, and summarizes the results mentioned above.  These graphs all
have distinct discrete rotational symmetries.  Obviously, there cannot
be a continuous transition from one discrete symmetry to another unless
one passes through the cylindrically symmetric triple point.  Thus, such
transitions are in general of first order.

\section{Exact treatment of the single-particle energies}
\label{sec:exact}

While our method treats the effect of the anharmonic term in the trapping
potential perturbatively, it can easily be extended to higher values
of $\lambda$ and/or $m_0$ by employing exact single-particle energies
and wave functions which include the effects of the anharmonicity to
all orders.

In Fig.\,5 we show the phase diagram that results from the
use of such exact solutions for the non-interacting system (i.e., by solving
the Schr\"odinger equation in the trapping potential $V(\rho)$ rather than
treating the anharmonic term perturbatively.)  The results are shown for
$\lambda = 0.05$ and $m_0 \leq 8$. In this case the phase boundary has no
simple scaling properties and depends on the specific choice of this parameter.
Although it is tempting to regard this value of $\lambda$ as small,
we find significant deviations from the results of Fig.\,2. [Note, however,
the quadratic dependence of the perturbative energy on $m$ in Eq.\,(\ref{sen}),
which implies that perturbation theory becomes less reliable with increasing $m$].
In this case the most unstable mode for $m_0 \leq 7$ is of the form 
$(m_1, m_0, m_2) = (0, m_0, 2m_0)$, while for $m_0 =8$ it is the one
with $(1, 8, 15)$. Therefore the triple point occurs 
\begin{figure}
\begin{center}
\epsfig{file=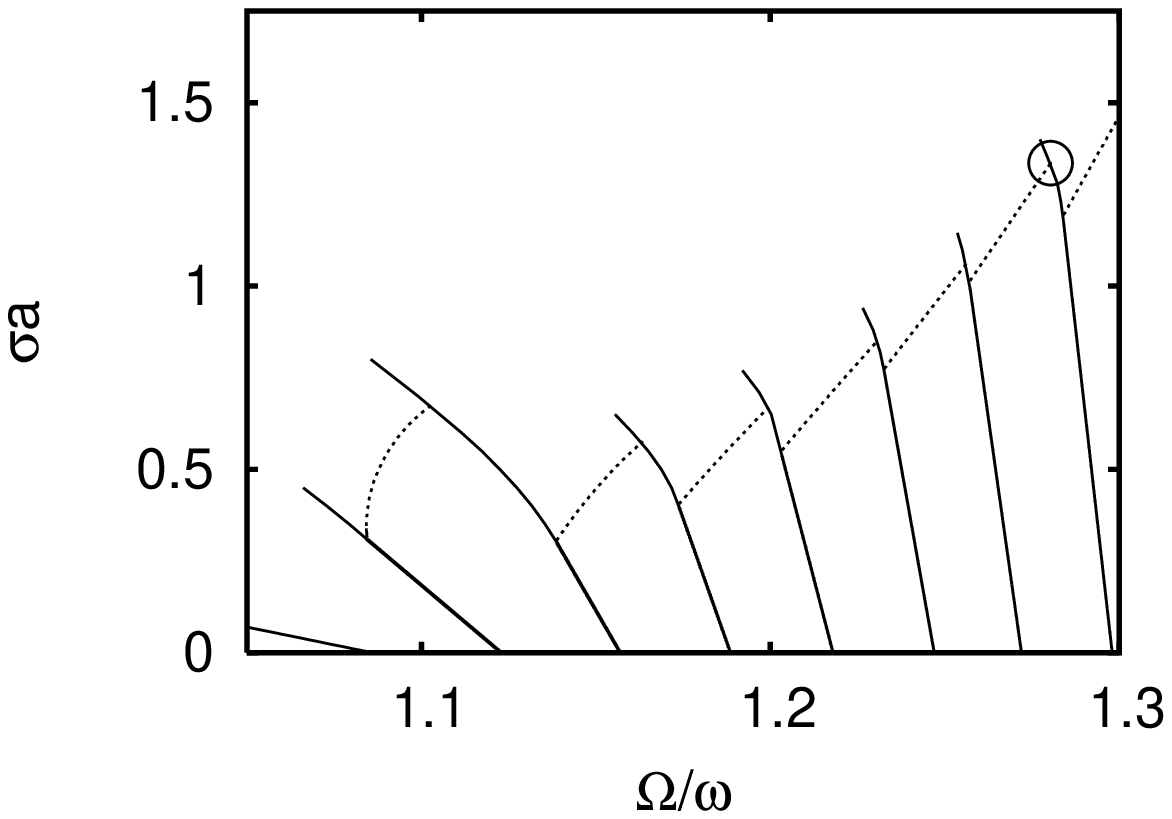,width=9.0cm,height=7.0cm,angle=0}
\vskip0.5pc
\begin{caption}
{The same graph as in Fig.\,2 for $0 \le m_0 \le 8$, with the single-particle
eigenvalue problem solved exactly. The circle denotes the triple point.}
\end{caption}
\end{center}
\label{FIG5}
\end{figure} 
\noindent
between the phases with $m_0 = 7$, $(m_1, m_0, m_2) = (0,7,14)$, and $(1,8,15)$, 
and is shown inside the circle in Fig.\,5.

\section{Numerical solutions of the Gross-Pitaevskii equation}
\label{sec:numerical}

In addition to the above variational calculations, we have also solved 
the Gross-Pitaevskii equation numerically. In order to locate the
continuous phase transition lines to a few digits precision, we have used
the following procedure. For given values of $\sigma a$ and $\Omega/\omega$, 
we take as initial state the harmonic-oscillator eigenfunction $\Phi_{m_0}$ 
with the addition of small-amplitude random noise in the core region.  
(The appropriate value of $m_0$ is easily determined numerically provided 
that we are not close to the first-order transition and is consistent with 
the variational results for the values reported here.)  This wave
function is propagated in imaginary time subject to the Gross-Pitaevskii 
equation in a rotating frame, and the overlaps, $c_{m'}$, of the wave function
with the harmonic-oscillator functions $\Phi_{m'}$ are monitored for a range 
of $m'$.  A steady increase in $|c_{m'}|$ for some $m'$ indicates that the 
multiply-quantized vortex state is energetically unstable towards a splitting 
into several vortices.  If, on the other hand, all $|c_{m'}|$ decrease with 
time, the multiply-quantized vortex is the energetically favorable 
configuration.  This procedure makes it easy to locate the position of 
the second-order transition without actually having to compute the ground 
state for each point in phase space. The latter procedure, while 
straightforward, is ill-suited to the problem at hand because convergence 
towards the ground state is extremely slow for the weak couplings considered
here.  In adition, the difference in energy between the many local energy 
minima is extremely small and requires exceedingly high precision.  In 
contrast, the method we have used is a fast and reliable way to locate 
the continuous phase transition lines and represents a significant extension 
of the variational study.

The results of this numerical study are shown in Fig.\,6. The
anharmonicity is chosen as $\lambda=0.005$.  According to the 
universality shown in Sec.\ \ref{sec:universality}, we need
only rescale the axes of Fig.\ 2 in order to obtain the
variational phase diagram for this new value of $\lambda$.
Agreement with the variational study is good for this very
weak anharmonicity, but the discrepancy grows rapidly with increasing 
$\lambda$.  Exact wave functions have been computed for a few values 
of $\sigma a$ and $\Omega/\omega$, and their symmetries are in agreement with 
those shown in Fig.\,4.

The first triple point occurs between $m_0=6$ and $m_0=7$, precisely 
as predicted by the variational method in Sec.\ \ref{sec:continuous}. 
The most unstable mode for $m_0=7$ thus involves the states with $(1,7,13)$. 
This is not inconsistent with the results reported in Sec.\ \ref{sec:exact}
computed using exact anharmonic-oscillator eigenfunctions, because the value of 
$\lambda$ was different in the two cases.  On the other hand, the present 
anharmonicity $\lambda=0.005$ is still large enough to result in one 
important difference compared to Figs.\ 2-4.  The triple point on the 
line of continuous transitions for $m_0=8$ is absent.  Thus, the most 
unstable mode is the one with $(m_1, m_0, m_2) = (1,8,15)$ for all values 
of $\Omega$ considered. We find instabilities of the form $(m_1,m_0,m_2)$ 
with $m_1\geq 2$ only for $m_0 \geq 9$.  This again emphasizes that very 
weak anharmonicities are required to ensure the quantitative validity of 
the variational approach.

\section{Summary}
\label{sec:conclusions}

In summary, we have studied a rotating Bose-Einstein condensate with
repulsive forces that is trapped in a quadratic-plus-quartic potential.
As the rotational frequency of the cloud and the coupling between the
atoms vary, the system exhibits three phases: a phase of multiple-quantization,
a phase of singly-quantized vortices, and a mixed phase. Our calculated
phase diagram turns out to be universal and partially exact in the
limit of weak coupling and small anharmonicity.

\vskip0.5pc
\noindent
The authors wish to thank Gordon Baym, Ben Mottelson, and
Chris Pethick for useful discussions. Author GMK acknowledges financial
support  from the Swedish Research Council (VR), and from the Swedish
Foundation for Strategic Research (SSF). Author EL acknowledges
financial support from the G{\"o}ran Gustafsson Foundation.

\begin{figure}
\begin{center}
\epsfig{file=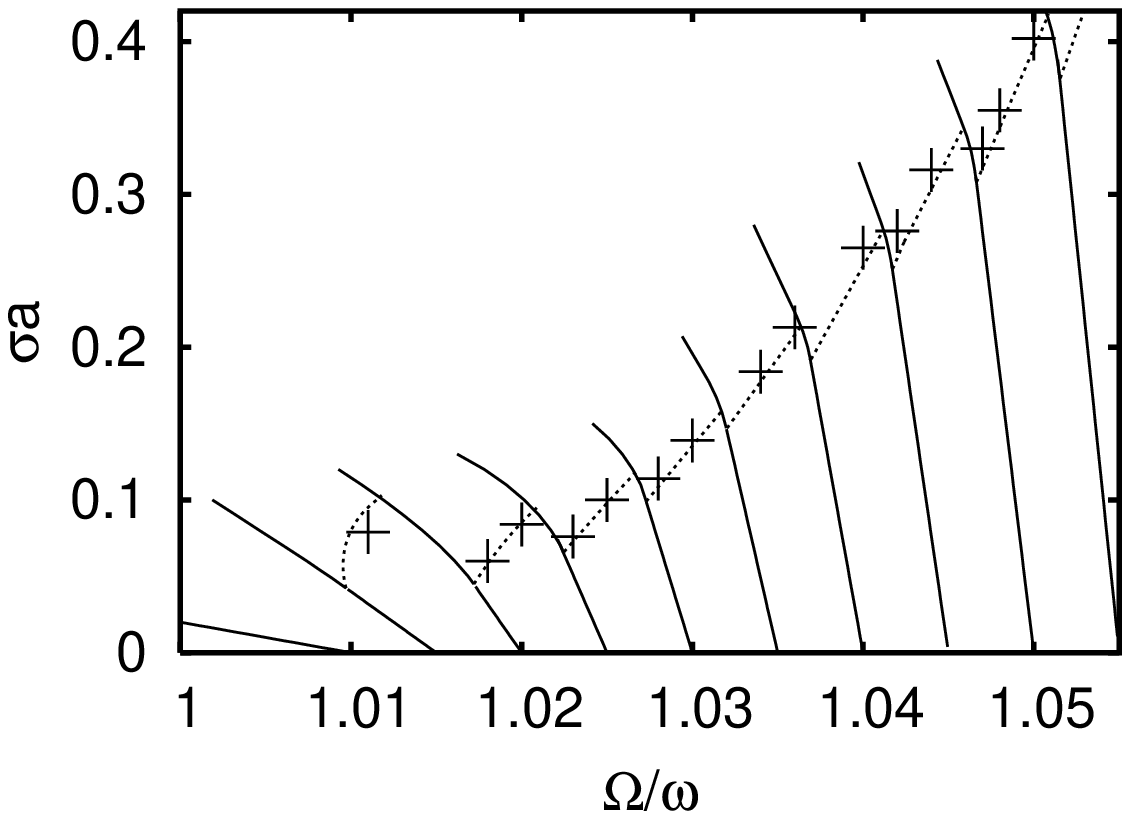,width=9.0cm,height=7.0cm,angle=0}
\vskip0.5pc
\begin{caption}
{Same graph as in Fig.\,2, for $\lambda=0.005$. The lines are identical
to those in Fig.\ 2, except that the axes have been scaled according to 
the universality of the phase diagram. The crosses represent 
points of the continuous transitions computed numerically from the
Gross-Pitaevskii equation (see Sec.\,\ref{sec:numerical}).}
\end{caption}
\end{center}
\label{FIG6}
\end{figure}

\end{document}